\documentclass[aps, prl, twocolumn, superscriptaddress, notitlepage,floatfix]{revtex4-2}
\usepackage{float}
\usepackage{bbm}
\usepackage{epsfig}
\usepackage{epstopdf}
\usepackage{graphicx}
\usepackage{amsmath,amssymb}
\usepackage{amsmath,bm}
\usepackage{amsmath,esint}
\usepackage{physics}
\usepackage{color}
\usepackage{hyperref}
\usepackage{url}
\usepackage{lineno,blindtext}
\usepackage[caption=false]{subfig}
\usepackage{siunitx, booktabs}
\usepackage{diagbox, eqparbox, hhline}
\usepackage{soul}
\usepackage{xcolor}
\usepackage[normalem]{ulem}
\newcolumntype{P}[1]{>{\centering\arraybackslash}p{#1}}

\begin{document}

\title{Emerging (2+1)D electrodynamics and topological instanton in pseudo-Hermitian two-level systems}

\author{Kuangyin Deng}
\email{kuangyid@ucr.edu}
\affiliation{Department of Electrical and Computer Engineering,  University of California, Riverside, California 92521, USA}
\author{Ran Cheng}
\email{rancheng@ucr.edu}
\affiliation{Department of Electrical and Computer Engineering,  University of California, Riverside, California 92521, USA}
\affiliation{Department of Physics and Astronomy, University of California, Riverside, California 92521, USA}
\affiliation{Department of Materials Science and Engineering, University of California, Riverside, California 92521, USA}

\begin{abstract}
We reveal a hidden electrodynamical structure emerging from a general $2\times2$ pseudo-Hermitian system that exhibits real spectra. Even when the Hamiltonian does not explicitly depend on time, the Berry curvature can be mapped onto a $2+1$ dimensional electromagnetic field arising from an artificial spacetime instanton, in sharp contrast to the Hermitian systems where the Berry curvature is equivalent to the static magnetic field of a magnetic monopole in three spatial dimensions. The instanton appearing as a spacetime singularity carries a topological charge that quantizes the jump of magnetic flux of the Berry curvature at the time origin. Our findings are demonstrated in a simple example related to antiferromagnetic magnons.
\end{abstract}

\maketitle

Geometric (Berry) phases and their topological origins entail profound implications in the physical behavior of quantum systems~\cite{xiaoRMP2010,provost1980riemannian,venuti2007,cheng2010quantum,asboth2016short}. In a general Hermitian quantum system, if the Hamiltonian depends on time implicitly through an adiabatic parameter $\bm{R}(t)$ but is not explicitly time-dependent, namely $H=H[\bm{R}(t)]$ rather than $H(\bm{R},t)$, the Berry phase is purely geometrical and can be viewed as the flux of a \textit{static} magnetic field (Berry curvature) residing in the parameter space spanned by $\bm{R}$, so long as $|d\bm{R}/dt|$ is much smaller than the energy gap---known as the adiabatic condition~\cite{xiaoRMP2010}. Unlike real magnetic fields that are always divergenceless, the Berry curvature diverges at the energy degenerate points which play the role of artificial magnetic monopoles~\cite{ArmitageRMP2018,yan2017topological}. The monopoles carry quantized magnetic charges that characterize the topological properties of the system.

However, for open and dissipative quantum systems that are typically described by non-Hermitian Hamiltonians, complex energy eigenvalues may arise, rendering the system unstable---either growing or decaying in time. The associated Berry phases could also become complex valued, compromising its geometrical interpretation. A salient exception is the pseudo-Hermitian (PH) systems, whose Hamiltonian satisfies
\begin{equation}
H^\dagger(\bm{R}) = \eta \, H(\bm{R}) \, \eta^{-1},
\label{eq:PH_condition}
\end{equation}
where $\eta$ is invertible, Hermitian, and independent of $\bm{R}$~\footnote{One can rescale $\eta$ to ensure $\eta^2=1$, by which $\det(\eta)$ cancels $\det(\eta^{-1})$. The Hermitian case is recovered if $\eta$ is the identity operator}. Under such a constraint, the eigenvalues of $H$ must appear in complex-conjugate pairs or simply be real-valued~\cite{mostafazadeh2002pseudo1,mostafazadeh2002pseudo2,ashida2020non,mostafazadeh2010pseudo}. Prior studies explored various scenarios of Berry phases in non-Hermitian systems~\cite{zhang2019time,garrison1988complex,liang2013topological,amaouche2022non}, and suggested that a real spectrum is crucial for defining a meaningful Berry phase~\cite{zhangPRA2019,cuiPRA2012,cheniti2020adiabatic}. Meanwhile, so long as real spectra are guaranteed~\footnote{The spectrum becomes purely real \textit{iff} there exists an invertible operator $\mathcal{O}$ such that $\eta=\mathcal{O}\mathcal{O}^\dagger$ in Eq.~\eqref{eq:PH_condition}. If such $\mathcal{O}$ does not exist, the eigenvalues will instead form
complex-conjugate pairs~\cite{mostafazadeh2002pseudo2}.}, all non-Hermitian systems fall into the same class as PH systems~\cite{mostafazadeh2002pseudo1,mostafazadeh2002pseudo2,ashida2020non}, which also includes the quasi-Hermitian cases~\cite{scholtz1992quasi}. Therefore, by focusing on the PH systems, one can effectively encompass all non-Hermitian models with real eigenvalues. Within the biorthonormal framework~\cite{brody2013,Kunst2018PRL}, a well-behaved and real Berry phase can be justified by the PH condition [see the Supplemental Material (SM)]. Nevertheless, what remains elusive is a magnetic interpretation, hence an intuitive analogy, of the Berry phases and their corresponding singularities serving as topological invariants.

In this Letter, we explore a generic $2\times2$ PH Hamiltonian exhibiting real spectra, which supports a unique Berry curvature that resembles a $2+1$ dimensional (D) electromagnetic field but cannot be simply mapped onto a static magnetic field as in its Hermitian counterpart, despite that the Hamiltonian $H(\bm{R})$ does not explicitly depend on time. The pivotal factor lies in an emergent \textit{spacetime} metric in the parameter space such that one component of $\bm{R}$ naturally plays the role of time while the rest act as space. We find that the electric and magnetic components of the Berry curvature are inherently connected by the $(2+1)$D Faraday equation in the presence of \textit{spacetime} singularities, dubbed instantons, carrying quantized topological charges. The topological instantons emerging in (real-spectral) PH systems parallel the role of magnetic monopoles appearing in Hermitian systems. This perspective unifies non-Hermitian Berry-phase physics with field-theoretic concepts, providing a novel handle on topology in open and dissipative quantum systems.

\textit{Emerging (2+1)D electrodynamics.}---Let us consider the most general two-level PH system, whose Hamiltonian depends on six real parameters~\cite{ashida2020non}
\begin{align}
H=a_0\sigma_0+(a_1\bm{n}_1+ia_2\bm{n}_2+ia_3\bm{n}_3)\cdot\bm{\sigma}, \label{eq:22Hamiltonian}
\end{align}
where $\bm{\sigma}=(\sigma_x,\sigma_y,\sigma_z)$ are the Pauli matrices, $\sigma_0$ is the identity matrix, while $\bm{n}_1=(\sin{\theta}\cos{\phi},\sin{\theta}\sin{\phi},\cos{\theta})$, $\bm{n}_2=(\cos{\theta}\cos{\phi},\cos{\theta}\sin{\phi},-\sin{\theta})$, and $\bm{n}_3=\bm{n}_1\times\bm{n}_2$, forming an orthonormal frame specified by $\theta$ and $\phi$. This Hamiltonian satisfies Eq.~\eqref{eq:PH_condition} with $\eta=\bm{n}_1\cdot\bm{\sigma}$. It is straightforward to solve the eigen-energies of $H$ as
\begin{align}
\varepsilon_\pm=a_0\pm\sqrt{a_1^2-a_2^2-a_3^2}=a_0\pm \abs{\ell},
\label{eq:NH-energy}
\end{align}
which are independent of $\theta$ and $\phi$ thanks to the $SO(3)$ symmetry among the base vectors $\bm{n}_i$ ($i=1,2,3$). This $SO(3)$ symmetry, owing to the form of coupling between $\bm{n}_i$ and $a_i$ in Eq.~\eqref{eq:NH-energy}, implies an $SO(1,2)$ symmetry among $a_1$, $a_2$ and $a_3$, which will become clear below. The exceptional points are located on the hypersurfaces of $a_1^2=a_2^2+a_3^2$, separating stable from unstable regions of the PH system. We will focus on the real-spectral cases satisfying $a_1^2>a_2^2+a_3^2$. In this region, the biorthonormal eigenstates~\cite{brody2013,Kunst2018PRL} can be solved as
\begin{subequations}
\label{eq:eigenstates}
\begin{align}
\ket{\psi^{R}_-}&=\frac{1}{2\ell}
\begin{bmatrix}
    -e^{i\phi}(a_1\sin\theta+ia_2\cos\theta+a_3) \\
    1+a_1\cos\theta-ia_2\sin\theta
\end{bmatrix},\\
\ket{\psi^{R}_+}&=\frac{1}{2\ell}
\begin{bmatrix}
    e^{-i\phi}(a_1\sin\theta+ia_2\cos\theta+a_3) \\
    1-a_1\cos\theta+ia_2\sin\theta
\end{bmatrix},\\
\ket{\psi^{L}_-}&=\frac{1}{\ell}
\left[
    \frac{e^{-i\phi}(-a_1\sin\theta+ia_2\cos\theta+a_3)}{1+a_1\cos\theta+ia_2\sin\theta},\ 1
\right]^T,\\
\ket{\psi^{L}_+}&=\frac{1}{\ell}
\left[
    \frac{e^{-i\phi}(-a_1\sin\theta+ia_2\cos\theta+a_3)}{-1+a_1\cos\theta+ia_2\sin\theta},\ 1
\right]^T,
\end{align}
\end{subequations}
where $R\,(L)$ represents the right (left) eigenstates and $\pm$ correspond to the $\pm$ energy level in Eq.~\eqref{eq:NH-energy}. Here we have assumed $\ell>0$, but none of the subsequent results will be altered if $\ell<0$.

Because $a_0$ represents an overall energy reference in Eq.~\eqref{eq:NH-energy}, which can be shifted to $0$, the parameter space is effectively 3D, spanned by $(a_1,a_2,a_3)$. Then a surface of constant energy $\varepsilon_0=\ell$ satisfies $\ell^2=a_1^2-a_2^2-a_3^2$, resembling a $(2+1)$D spacetime interval in special relativity (with the speed of light $c=1$). Based on this analogy, it is tempting to identify a coordinate mapping:
\begin{align}
 a_1\rightarrow t,\quad a_2\rightarrow x,\quad a_3\rightarrow y,
 \label{eq:mapping}
\end{align}
by which we could interpret the parameter space as an artificial \textit{spacetime} endowed with a Minkowski metric $\eta_{\mu\nu}=\text{diag}(1,-1,-1)$ so that $\ell^2=t^2-x^2-y^2$. Let $r^\mu$ enumerating through $(t,x,y)$, then $\ell^2=\eta_{\mu\nu}r^\mu r^\nu$. This intriguing structure implies that a $2\times2$ PH Hamiltonian is intimately related to an emergent spacetime covariance, which can be attributed to the $SO(3)$ symmetry of the basis $\bm{n}_i$ under the PH condition in Eq.~\eqref{eq:PH_condition}.

\begin{figure}[t]
\centering
\includegraphics[width=0.9\linewidth]{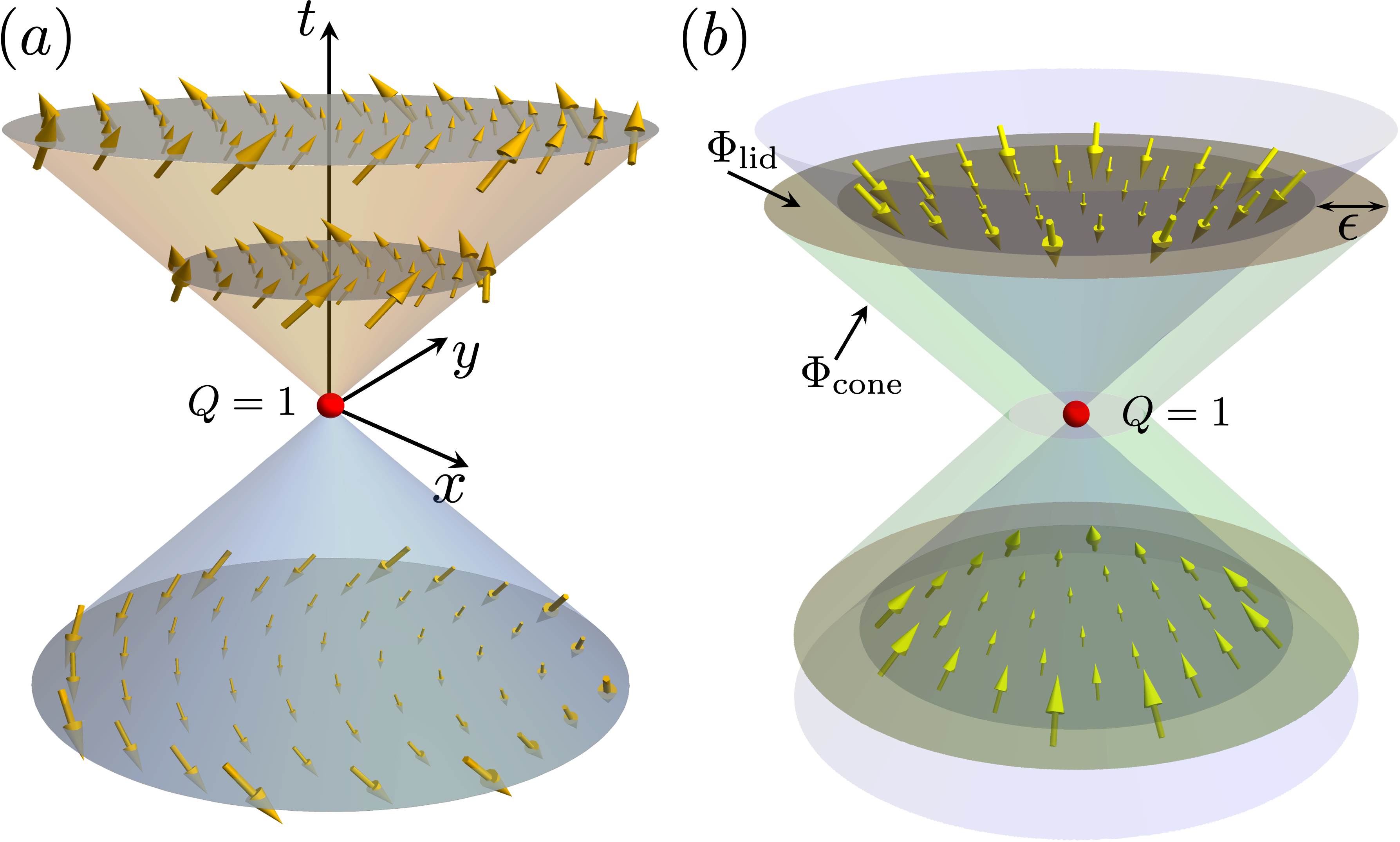}
\caption{(a) Vector plot of the emerging gauge fields $(B,E_x,E_y)$ with its norm scaled logarithmically, on three parallel cutting planes within the light cone. The instanton ($Q=1$ for $\varepsilon_-$) is indicated by a red dot. (b) Gauss' surface enclosing the instanton, with the dual field vector $\tilde{\bm{F}}$ plotted on two cutting planes for positive and negative $t$.}
\label{fig:fields}
\end{figure}

Using the biorthonormal eigenstates, we can define the Berry connection as ($\hbar=1$ in natural units)
\begin{align}
A_{\mu}^{\pm}=i\bra{\psi^L_\pm}\partial_\mu\ket{\psi^R_\pm}
\label{eq:connection}
\end{align}
with $\partial_\mu\equiv\partial/\partial r^\mu$. Accordingly, we can define the Berry curvature as
\begin{align}
F_{\mu\nu}^{\pm}=\partial_\mu A_{\nu}^{\pm}-\partial_\nu A_{\mu}^{\pm},
\label{eq:em tensor}
\end{align}
which is antisymmetric $F_{\mu\nu}^{\pm}=-F_{\nu\mu}^{\pm}$. Since $F_{\mu\nu}^+=-F_{\mu\nu}^-$ (\textit{i.e.}, the curvatures associated with the two energy levels always cancel), we can simplify the notations by focusing on $F_{\mu\nu}^-$ and omitting the $\pm$ index, unless it is necessary to specify the energy level. In the matrix form, the Berry curvature is expressed as
\begin{align}
F_{\mu\nu}=\frac{-\epsilon_{\mu\nu\rho}r^{\rho}}{2\ell^3}
=\frac{1}{2(t^2-x^2-y^2)^{3/2}}
\begin{pmatrix}
0 & -y & x\\
y & 0 & -t\\
-x & t & 0
\end{pmatrix},
\label{eq:Ftensor}
\end{align}
where $\ell=(\eta_{\mu\nu}r^\mu r^\nu)^{1/2}=\sqrt{t^2-x^2-y^2}$ and $\epsilon_{\mu\nu\rho}$ is the Levi-Civita symbol. The gauge field is $U(1)$ invariant because $\varepsilon_{\pm}$ is non-degenerate for $\ell\neq0$. In terms of the electromagnetic field variables, Eq.~\eqref{eq:Ftensor} translates into
\begin{align}
B=\frac{t}{2\ell^3},\quad
E_x=&-\frac{y}{2\ell^3},\quad
E_y=\frac{x}{2\ell^3},
\label{eq:em fields}
\end{align}
which is illustrated in Fig.~\ref{fig:fields}(a). By sharp contrast, in a Hermitian setting, all coordinates of the parameter space are akin to pure spatial variables so that the Berry curvature can only be mapped onto a static magnetic field rather than an electromagnetic tensor. A simple example is $H=\sum_kx_k\sigma _k$ with $x_k=x,y,z$, whose eigen-energies are $\varepsilon_{\pm}=\pm r$ with $r^2=\sum_kx_k^2$, and the associated Berry curvature $B_k=\mp x_k/2r^3$ is a static magnetic field.

\textit{Spacetime instanton.}---The Bianchi identity requires that $\epsilon^{\rho\mu\nu}\partial_{\rho}F_{\mu\nu}=0$, corresponding to Faraday's equation $\partial_tB+\hat{\bm{z}}\cdot(\bm{\nabla}_\perp\!\times\bm{E})=0$ where $\bm{E}=(E_x,E_y,0)$ and $\bm{\nabla}_\perp=(\partial_x,\partial_y,0)$ is the planar nabla operator. A naïve substitution of Eq.~\eqref{eq:em fields} indeed satisfies the above relation, but there is a caveat. Recall that a point charge represented by $\delta(r)$ generates an electric field proportional to $1/r^2$, which, upon direct application of the Gauss-Coulomb law, yields zero and cannot reproduce $\delta(r)$~\cite{Griffiths}. This is because the spatial derivatives become ill-defined at $r\rightarrow0$ while $\delta(r)$ is non-analytical. By the same token, properly treating the diverging behavior of Eq.~\eqref{eq:em fields} for $\ell\rightarrow0$ leads to a modified equation:
\begin{align}
\partial_tB+\hat{\bm{z}}\cdot(\bm{\nabla}_\perp\!\times\bm{E})=-2\pi Q\delta(t)\delta(x)\delta(y),
\label{eq:Faraday}
\end{align}
where $Q=1$ is the topological charge of an \textit{instanton} located at the spacetime origin~\cite{Pisarski1986,henneaux1986quantization}, at which the energies become degenerate. It is easy to see that $Q=\mp1$ for the $\varepsilon_{\pm}$ level, respectively. We stress that the \textit{instanton} here refers to a pointlike instantaneous singularity in the (2+1)D Minkowski space.

To better appreciate the appearance of instanton as the \textit{source} of Berry curvature, we turn to the \textit{dual} field $\Tilde{F}^{\mu}=\epsilon^{\mu\alpha\beta}F_{\alpha\beta}$, in terms of which the Bianchi identity becomes $\partial_\mu\tilde{F}^\mu=0$ away from the spacetime origin. If we regard $\Tilde{F}^{\mu}$ as a 3D vector $\tilde{\bm{F}}=(-B,-E_y,E_x)$ and define the 3D nabla operator $\tilde{\bm{\nabla}}=(\partial_t,\partial_x,\partial_y)$, then the Bianchi identity is further transformed into $\tilde{\bm{\nabla}}\cdot\tilde{\bm{F}}=0$, which enables us to intuitively reinterpret Eq.~\eqref{eq:Faraday} as Gauss's law in 3D:
\begin{align}
\tilde{\bm{\nabla}}\cdot\tilde{\bm{F}}=2\pi Q\delta^3(\tilde{\bm{r}}),
\label{eq:dualfield}
\end{align}
where $\tilde{\bm{r}}=(t,x,y)$ is identified as $r^{\mu}$.

It is worthwhile emphasizing that magnetic monopoles are conceptually incompatible with the framework of $(2+1)$D Maxwell's theory in which the magnetic field $B$ is a pseudo-scalar so that the magnetic Gauss's law is meaningless; magnetic monopoles must yield to instantons. Nevertheless, the Faraday equation in the presence of a spacetime instanton can be intuitively placed in a hypothetical $(3+1)$D context through dimensional reduction. Specifically, if magnetic current density $\bm{J}_m$ exists, the $(3+1)$D Faraday's equation reads $\bm{\nabla}\times\bm{E}+\partial_t\bm{B}=-\bm{J}_m$. When everything becomes uniform in the $z$ direction, and $\bm{J}_m=2\pi Q\delta(t)\delta(x)\delta(y)\hat{\bm{z}}$ represents an infinitely narrow pulse magnetic current at $t=0$, Eq.\eqref{eq:Faraday} is retrieved. In this regard, the instanton charge $Q$ is nothing but the amplitude of the pulse stimulus.

To further demystify the topological implication of the spacetime instanton, we consider a Gauss surface enclosing the instanton as illustrated in Fig.~\ref{fig:fields}(b), which consists of two reversely oriented cones slightly detached from the lightcone by $\epsilon$~\footnote{Outside of the lightcone, the eigen-energies become imaginary, but the eigenstates adopt the same form as in Eq.~\eqref{eq:eigenstates}, hence the Berry curvature retains the same form.}. After some tricky algebra detailed in the SM, we can associate the total flux of the dual field $\tilde{\bm{F}}$ with the topological charge $Q$ by virtue of the integral form of Eq.~\eqref{eq:dualfield}: $\oiint\tilde{\bm{F}}\cdot d\bm{s}=2\pi Q$. This relation differs from the ordinary Gauss theorem in 3D Euclidean space by a factor of $2$. In terms of the original field variables, the Gauss relation becomes
\begin{align}
\Phi(t_0)-\Phi(-t_0)+\int_{-t_0}^{t_0}dt\oint_{C(t)}\bm{E}\cdot d\bm{l}=-2\pi Q, \label{eq:interpret}
\end{align}
where $\pm t_0$ parameterize the top and bottom lids of the Gauss surface and $\Phi(t)=\iint\bm{B}(t)\cdot d\bm{s}$ is the magnetic flux through the surface bounded by circle $C(t)$ at time $t$ (for $-t_0\le t\le t_0$). In the $t_0\rightarrow0^+$ limit, $\Phi(t_0)-\Phi(-t_0)$ becomes $\Phi(0^+)-\Phi(0^-)$, \textit{i.e.}, the jump of magnetic flux $\Delta\Phi(0)$ at $t=0$, while $\oint_{C(0)}\bm{E}\cdot d\bm{l}$ (the vorticity of the $\bm{E}$ field) is continuous across $t=0$ so that the third term in Eq.~\eqref{eq:interpret} vanishes for $t_0\rightarrow0^+$, whereas $-2\pi Q$ on the right-hand side is not affected because the instanton is still enclosed by the Gauss surface. Therefore, Eq.~\eqref{eq:interpret} indicates $\Delta\Phi(0)/2\pi=-Q$, entailing a neat topological interpretation: the discontinuity of magnetic flux at $t=0$ is quantized in integer multiples of $2\pi$, with the integer being the instanton number.

In a material context, the artificial spacetime variables $(t,x,y)$ are typically functions of the 3D crystal momentum $\bm{k}$, and the spacetime instanton corresponds to the solution of $t(\bm{k})=x(\bm{k})=y(\bm{k})=0$. In comparison, a well-known material context for the $3$D magnetic monopoles associated with Hermitian systems is the $3$D Weyl semimetals~\cite{ArmitageRMP2018}. Analogous to the stability of the Weyl points, here the three equations $r^\mu(\bm{k})=0$ define three distinct $2$D surfaces in the $\bm{k}$ space, and their intersection (if exists) typically reduces to a point being robust against small deformations of $r^\mu(\bm{k})$, thus rendering the instanton topologically stable. In a PH lattice model, the total number of instantons is a topological invariant of the parameter space, obtained by integrating over the region enclosed by the Fermi surface, provided that the eigen-energies remain real. In particular, if $(k_x, k_y, k_z)$ can be linearly mapped onto $(t,x,y)$ (e.g., by Taylor expansion around a high symmetry point), then the gauge-field interpretation conceived above will be directly applicable to the $\bm{k}$ space.

\textit{Field dynamics.}---Concerning the emerging $(2+1)$D electrodynamics, it is instructive to check the
self-dynamics of the gauge field consisting of Coulomb's law and Ampere's law. In this regard, we find that Eq.~\eqref{eq:Ftensor} satisfies $J_\nu=\partial^\mu F_{\mu\nu}=0$, or
\begin{align}
 \bm{\nabla}_\perp\!\cdot\bm{E}=0,\qquad \partial_t\bm{E}-\bm{\nabla}_\perp\!\times(B\hat{\bm{z}})=0,
 \label{eq:Ampere}
\end{align}
meaning that both the electric charge density $J_0$ and the electric current density $J_i$ vanish exactly. Equations~\eqref{eq:Faraday} and~\eqref{eq:Ampere} together indicate that the Berry curvature emerging in our $2\times2$ PH system can be described by an effective Maxwell Lagrangian density $\mathcal{L}=-F_{\mu\nu}F^{\mu\nu}$ devoid of electric sources, while the functional derivative $\delta\mathcal{L}/\delta A_\mu=0$ is subject to the topological constraint in the form of a spacetime instanton.

With a full set of Maxwell's equations, our Berry curvature admits a compelling interpretation through the lens of $(2+1)$D electrodynamics. At $r^\mu=(0,0,0)$, an instanton flashes infinitely fast, triggering an electromagnetic wave propagating in spacetime. This scenario contrasts sharply with what one normally expects for an Hermitian system, where a magnetic monopole persists indefinitely and the gauge field it generates is not self-dynamical. Moreover, even when an Hermitian Hamiltonian explicitly depends on time, $H=H[\bm{R}(t),t]$ (but $\partial_tH$ is small enough to maintain adiabaticity), which affords both electric and magnetic components in the Berry curvature, only Faraday's law can be satisfied whereas Coulomb's law and Ampere's law are baseless~\cite{niu2017physical}. Without acquiring any self-dynamics, the gauge field emerging in an Hermitian system can only passively affect the motion of particles carrying gauge charges but cannot be inversely excited by the particles.

\textit{Quantum geometric tensor.}---In Hermitian systems, the Berry curvature can be viewed as the anti-symmetric component of a broader geometric concept known as the quantum geometric tensor (QGT)~\cite{QGT}, whose symmetric component is the quantum metric. Recently, the QGT has been generalized into non-Hermitian systems~\cite{zhang2019quantum,zhu2021band,QGTnonH,behrends2025quantum}, but little is known about its manifestation in PH systems that host topological instantons. Based on Eq.~\eqref{eq:PH_condition}, we can derive the QGT associated with a particular energy level $n$ for a general PH Hamiltonian as
\begin{align}
Q_{n,\mu\nu}=&\bra{\partial_\mu\psi_n^L}(1-P)\ket{\partial_\nu\psi_n^R},
\end{align}
where $P=\ket{\psi_n^R}\bra{\psi_n^L}$ is the projection operator. By exploiting the biorthonormal Hellmann–Feynman relations, we can express the QGT in terms of the eigenstates and the Hamiltonian:
\begin{align}
Q_{n,\mu\nu}=&\sum_{m\ne n}\frac{\bra{\psi_n^L}\partial_\mu H\ket{\psi_m^R}\bra{\psi_m^L}\partial_\nu H\ket{\psi_n^R}}{(E_n-E_m)^2},
\label{eq:QGT}
\end{align}
which looks similar to its Hermitian counterpart but the left and right eigenvectors are distinguishable. We find that away from the energy degenerate points, the QGT can be decomposed as
\begin{align}
Q_{n,\mu\nu}=&g_{n,\mu\nu}-\frac{i}{2}F_{n,\mu\nu},
\end{align}
where $g_{\mu\nu}$ is the quantum metric (a symmetric tensor) and the $F_{\mu\nu}$ is the Berry curvature we defined previously. This relation is identical to its Hermitian counterpart~\cite{cheng2010quantum}. By inserting Eq.~\eqref{eq:eigenstates} into Eq.~\eqref{eq:QGT} and taking its symmetric component, we finally obtain the quantum metric associated with the spacetime instanton (with $Q=1$) in a general $2\times2$ PH system:
\begin{align}
&g_{\mu\nu}=\frac{1}{4\ell^4}\left[\ell^2\eta_{\mu\nu}-(\eta_{\mu\alpha}r^\alpha)\otimes(\eta_{\nu\beta}r^\beta)\right] \notag\\
&=\frac{-1}{4(t^2-x^2-y^2)^2}\begin{pmatrix}
x^2+y^2 & -tx & -ty \\[1ex]
-tx & t^2-y^2 & xy \\[1ex]
-ty & xy & t^2-x^2
\end{pmatrix},
\end{align}
which is related to the Berry curvature by
\begin{align}
 \frac14 F_{\mu\nu}F_{\alpha\beta}=g_{\mu\alpha}g_{\nu\beta}-g_{\mu\beta}g_{\nu\alpha}.
\end{align}

\textit{Example.}---Let us consider a concrete $2\times 2$ PH Hamiltonian that can effectively describe topological magnons in transition-metal trichalcogenides with collinear antiferromagnetic order on a honeycomb lattice~\cite{ran2016,zhang2022perspective}. In this setting, magnons of opposite spin species form a two-level system governed by a Schrödinger-like equation $i\hbar\sigma_z \partial_\tau \Psi=H_0\Psi$, where $H_0=tI+x\sigma_x+y\sigma_y+a_0\sigma_z$ with $\sigma_i$ being the Pauli matrices. The appearance of $\sigma_z$ on the left hand side is attributed to the bosonic Bogoliubov transformation. Multiplication of $\sigma_z$ on both sides yields a true PH system satisfying
\begin{align}
 i\hbar\partial_\tau\Psi=H_{\rm ph}\Psi,
\end{align}
where the Hamiltonian
\begin{align}
 H_{\rm ph}=\sigma_zH_0=t\sigma_z+ix\sigma_y-iy\sigma_x+a_0I
\end{align}
is a particular case of Eq.~\eqref{eq:22Hamiltonian} for $\theta=0$, $\phi=\pi/2$, $\bm{n}_1=\hat{\bm{z}}$, $\bm{n}_2=\hat{\bm{y}}$ and $\bm{n}_3=-\hat{\bm{x}}$. Basing on the form of Eq.~\eqref{eq:NH-energy}, we can adopt a hyperbolic parameterization
\begin{align}
t=\ell\cosh{\xi},\quad x=\ell\sinh{\xi}\cos{\lambda},\quad  y=\ell\sinh{\xi}\sin{\lambda},
\end{align}
which reduces the eigenstates in Eq.~\eqref{eq:eigenstates} into
\begin{align}
\ket{\psi^{R}_-}=&
\begin{pmatrix}
-e^{-i\lambda}\sinh{\frac{\xi}{2}}\\
\cosh{\frac{\xi}{2}}
\end{pmatrix},\,
\ket{\psi^{R}_+}=
\begin{pmatrix}
e^{-i\lambda}\cosh{\frac{\xi}{2}}\\
-\sinh{\frac{\xi}{2}}
\end{pmatrix},\nonumber\\
\ket{\psi^{L}_-}=&
\begin{pmatrix}
e^{-i\lambda}\sinh{\frac{\xi}{2}}\\
\cosh{\frac{\xi}{2}}
\end{pmatrix},\quad
\ket{\psi^{L}_+}=
\begin{pmatrix}
e^{-i\lambda}\cosh{\frac{\xi}{2}}\\
\sinh{\frac{\xi}{2}}
\end{pmatrix},
\end{align}
where they are all independent of $\ell$. Following a straightforward algebra elaborated in the SM, we can obtain the QGT under the $(t,\,x,\,y)$-basis and fully recover the emerging electromagnetic field in Eq.~\eqref{eq:Ftensor} featuring an instanton solution at the origin.

What is interesting here is that the QGT under the $(\ell,\,\xi,\,\lambda)$-basis becomes block diagonal with a rank equal to $2$, as all elements involving $\ell$ vanish (see details in the SM). This means that the effective dimension of the parameter space is reduced to $2$. The Berry curvature and the quantum metric residing in the $2$D space spanned by $\xi$ and $\lambda$ (for the lower band $\varepsilon_-$) become:
\begin{align}
F_{ij}&=\frac12
\begin{pmatrix}
0 & -\sinh{\xi} \\
\sinh{\xi} & 0
\end{pmatrix}, \\
g_{ij}&=\frac{-1}4
\begin{pmatrix}
1 & 0 \\
0 & \sinh^2{\xi}
\end{pmatrix},
\end{align}
which characterize the Hilbert space geometry on an equal energy surface (labeled by a non-zero constant $\ell$). For the upper band, $F_{ij}$ flips sign while $g_{ij}$ remains the same. In this $2$D parameter space, the Berry curvature has only one component, namely the off-diagonal element $f=(1/2)\sinh\xi$, satisfying
\begin{align}
f=2\sqrt{\text{det}[g]},
\end{align}
which coincides with a well-known identify in $2\times2$ Hermitian systems~\cite{kolodrubetz2017geometry}.

\textit{Final remarks.}---It is instrumental to mention that the $2\times2$ $\mathcal{PT}$-symmetric systems form a subclass of the general PH systems we have considered. For instance, a $\mathcal{PT}$-symmetric Hamiltonian $H_{\rm pt}=a_0I+ix_0\sigma_x+y_0\sigma_y+z_0\sigma_z$ satisfying $\sigma_zH^*_{\rm pt}\sigma_z^{-1}=H_{\rm pt}$ supports real eigenvalues
\begin{align}
 \varepsilon_{\pm}=a_0\pm\sqrt{-x_0^2+y_0^2+z_0^2},
\end{align}
which can be mapped onto Eq.~\eqref{eq:NH-energy} by identifying $x_0\rightarrow-a_3$, $y_0\rightarrow a_1\sin\theta$, and $z_0\rightarrow a_1\cos\theta$, while $a_2=0$ and $\theta$ is arbitrary. Under such re-parameterization, when setting $\phi=\pi/2$, Eq.~\eqref{eq:22Hamiltonian} directly constructs our chosen $H_{\rm pt}$. Therefore, a $\mathcal{PT}$-symmetric system with purely real spectra is fully encompassed by our model.

One may wonder if there exists a mapping from a PH Hamiltonian to the $(3+1)$D electrodynamics with a full set of Maxwell equations. While Eq.~\eqref{eq:22Hamiltonian} represents the most general $2\times2$ PH models, its spectrum depends on only three independent variables (apart from $a_0$---an overall energy shift), which is insufficient to account for the six components of the $(3+1)$D electromagnetic field tensor. On the other hand, a $3\times3$ PH Hamiltonian necessarily involves the eight Gell-Mann matrices and the identity, holding more degrees of freedom than what are needed to describe the $(3+1)$D electrodynamics. This discrepancy poses an open question for future explorations.

\begin{acknowledgments}
K.D. would like to thank Jie Zhou for helpful discussions. This work is supported by the UC Regents' Faculty Development Award, and the National Science Foundation under Award No. DMR-2339315.
\end{acknowledgments}

\clearpage
\onecolumngrid
\begin{center}
  \textbf{\large Supplementary Material}\label{supp}
\end{center}

\setcounter{section}{0}
\setcounter{subsection}{0}
\setcounter{subsubsection}{0}
\setcounter{secnumdepth}{3}

\renewcommand{\thesection}{S\arabic{section}}
\renewcommand{\thesubsection}{S\arabic{section}.\arabic{subsection}}
\renewcommand{\thesubsubsection}{S\arabic{section}.\arabic{subsection}.\arabic{subsubsection}}

\setcounter{equation}{0}
\renewcommand{\theequation}{S.\arabic{equation}}

\section{Berry phase is real for a general PH Hamiltonian with real spectra}\label{supsec:1}

Consider a general non-Hermitian Hamiltonian undergoing an adiabatic evolution in time. The time-dependent eigenstate at time $t$ is expressed by
\begin{align}
\ket{\psi^R_n(t)}=e^{-i\theta(t)}\ket{n^R(\bm{R}(t))},
\end{align}
where $\bm{R}(t)$ is the adiabatic parameter, $\ket{n^R(\bm{R}(t))}$ is an instantaneous right eigenvector with eigenenergy $E_n(\bm{R})$ such that $H(\bm{R})\ket{n^R(\bm{R})}=E_n(\bm{R})\ket{n^R(\bm{R})}$, and $\theta$ includes both the dynamical and the Berry phases. To satisfy the biorthonormality condition
\begin{align}
 \bra{\psi^L_n}\ket{\psi^R_n}=\bra{n^L(\bm{R})}\ket{n^R(\bm{R})}=1,
\end{align}
the corresponding left eigenvector must be chosen as $\bra{\psi^L_n(t)}=e^{i\theta(t)}\bra{n^L(\bm{R})}$. Accordingly, the total phase is given by
\begin{align}
\theta (t)=&\frac{1}{\hbar}\int^t_0 E_n(\bm{R})dt'-i\int^t_0\bra{n^L(\bm{R})}\ket{\partial_t n^R(\bm{R})}dt'.\label{eq:non-hermitian phase}
\end{align}
In the most general scenario, both the first term (dynamical phase) and the second term (Berry phase) may be complex-valued, leading to either decaying or growing time evolution of the quantum state. Now, let us impose the PH condition
\begin{align}
H^\dagger(\bm{R})=\eta H(\bm{R})\eta^{-1},\label{eq:ph-condition}
\end{align}
where $\eta$ is invertible, Hermitian, and independent of $\bm{R}$. Under the above condition, the eigenenergies can either be real or come in conjugate pairs (see Refs.~[9,\,10,\,12] of the main text). For our purposes, we only focus on the regime of real eigenenergies. By employing the PH condition, we have
\begin{align}
H^\dagger\eta\ket{n^R}=\eta H\ket{n^R}=E_n\eta\ket{n^R}=E_n^*\left(\eta\ket{n^R}\right),
\end{align}
so $\eta\ket{n^R}$ must be a left eigenvector with eigenenergy $E_n^*$, namely $\ket{n^L}=\lambda_n\eta \ket{n^R}$. Because of the biorthonormal condition $\bra{n^L}\ket{n^R}=\lambda_n^*\bra{n^R}\eta\ket{n^R}=1$ and that $\bra{n^R}\eta\ket{n^R}$ is real, the constant $\lambda_n$ must be real. Moreover,
\begin{align}
0=\partial_t \langle n^L(\bm{R})\ket{n^R(\bm{R})}&=\partial_t [\lambda_n\bra{n^R(\bm{R})}\eta\ket{n^R(\bm{R})}]\nonumber\\
&=\lambda_n\bigg[\bra{n^R(\bm{R})}\eta\ket{\partial_t n^R(\bm{R})}+\bra{\partial_t n^R(\bm{R})}\eta\ket{n^R(\bm{R})}\bigg].
\end{align}
Since the two terms in the bracket are complex conjugates of each other, they must be purely imaginary when $\lambda_n$ is non-zero. This ensures that the Berry-phase term in Eq.~\eqref{eq:non-hermitian phase} is real.\\

\section{Flux of the Dual Field $\tilde{\bm{F}}$ involving \(\Phi_{\rm cone}\) and \(\Phi_{\rm lid}\)}

As shown in Fig.~1(b) of the main text, the Gauss surface is slightly detached from the light cone in order to fully enclose the instanton. As the speed of light is $c=1$, we can regulate the side surface of the upper cone as
\begin{align}
 S_{\rm cone}:\quad t=\sqrt{x^2+y^2}-\epsilon,\quad 0\le t\le1,
\end{align}
and correspondingly the upper lid is described by
\begin{align}
 S_{\rm lid}:\quad x^2+y^2=(1+\epsilon)^2\ \mbox{and}\ t=1.
\end{align}
We parameterize the $S_{\rm cone}$ as
\begin{align}
t=r-\epsilon,\quad x=r\cos\theta,\quad y=r\sin\theta,\quad 0\le r\le r_{\rm max}\,,\quad 0\le \theta<2\pi\,,
\end{align}
so $r_{\rm max}=1+\epsilon$ at the top ($t=1$) and $r_{\rm min}=\epsilon$ at the bottom ($t=0$), and
\begin{align}
\ell=\sqrt{t^2-r^2}=\sqrt{\epsilon^2-2\epsilon r}=i\sqrt{2\epsilon r-\epsilon^2}.
\end{align}
Since $S_{\rm cone}$ is outside the light cone, the dual field becomes imaginary
\begin{align}
\tilde{\bm{F}}=-\frac{i}{2(2\epsilon r-\epsilon^2)^\frac{3}{2}}\Bigl(r-\epsilon,\,r\cos\theta,\,r\sin\theta\Bigr).
\end{align}

To perform the surface integral, we need to determine the surface element on $S_{\rm cone}$. Under the $(r,\theta)$ parameterization, the position vector reads
\begin{align}
\bm{r}(r,\theta)=\Bigl(r-\epsilon,\,r\cos\theta,\,r\sin\theta\Bigr),
\end{align}
whose directional derivatives are $\bm{r}_r=\left(1,\,\cos\theta,\,\sin\theta\right)$ and $
\bm{r}_\theta=\left(0,\,-r\sin\theta,\,r\cos\theta\right)$. Their cross product is
\begin{align}
\bm{r}_r\times\bm{r}_\theta
=\left(r,\; -r\cos\theta,\; -r\sin\theta\right)
\end{align}
with its magnitude being $|\bm{r}_r\times\bm{r}_\theta|=\sqrt{2}r$, thus the normal vector is
\begin{align}
 \hat{\bm{n}}=-\frac{\bm{r}_r\times\bm{r}_\theta}{|\bm{r}_r\times\bm{r}_\theta|},
\end{align}
where the minus sign ensures that $\hat{\bm{n}}$ is pointing outward. Correspondingly, the area element is
\begin{align}
d\bm{s}=\hat{\bm{n}}ds=\hat{\bm{n}}\sqrt{2}r\,dr\,d\theta\,.
\end{align}
Now, the dot product between $\tilde{\bm{F}}$ and $\hat{\bm{n}}$ is
\begin{align}
\tilde{\bm{F}}\cdot\hat{\bm{n}}
&=\frac{i}{2(2\epsilon r-\epsilon^2)^\frac{3}{2}}\frac{1}{\sqrt{2}r}\Bigl(r-\epsilon,\,r\cos\theta,\,r\sin\theta\Bigr)\cdot\Bigl(r,\; -r\cos\theta,\; -r\sin\theta\Bigl) \nonumber\\
&=\frac{-i}{8\epsilon^{\frac{1}{2}}(r-\epsilon/2)^\frac{3}{2}},
\end{align}
so the flux through the (upper) $S_{\rm cone}$ is
\begin{align}
\Phi_{\rm cone}=\iint d\bm{s}\cdot\tilde{\bm{F}}
=\int_{0}^{2\pi}\!d\theta\int_{r_{\rm min}}^{r_{\rm max}}
\frac{-i\sqrt{2}rdr}{8\epsilon^{\frac{1}{2}}(r-\epsilon/2)^\frac{3}{2}}\approx -2\pi\frac{\sqrt{2}i}{4\epsilon^{\frac{1}{2}}}=-\frac{\pi i}{\sqrt{2\epsilon}},
\end{align}
where in the last step we have taken $\epsilon\rightarrow0$.

For the flux on the upper lid $\Phi_{\rm lid}$, we parameterize $S_{\rm lid}$ using $(r,\theta)$ again such that
\begin{align}
x=r\cos\theta,\quad y=r\sin\theta,\quad 0\le r\le r_{\rm max}\,,\quad 0\le \theta<2\pi\,,
\end{align}
where $r_{\rm max}=t+\epsilon$. Because the surface normal is simply $\hat{\bm{t}}$, we have
\begin{align}
 \tilde{\bm{F}}\cdot\hat{\bm{n}}=\tilde{F}^t=-\frac{1}{2(1-r^2)^\frac{3}{2}},
\end{align}
so the flux through the lid is
\begin{align}
\Phi_{\rm lid}&=-\int_S\frac{dxdy}{2(1-x^2-y^2)^{\frac{3}{2}}}=-\int_0^{1+\epsilon}\int_0^{2\pi}\frac{rdrd\theta}{2(1-r^2)^{\frac{3}{2}}}\nonumber\\
&=\frac{\pi}{2}\int_0^{1+\epsilon}\frac{d(1-r^2)}{(1-r^2)^{\frac{3}{2}}}=-\frac{\pi}{\sqrt{-2\epsilon-\epsilon^2}}+\pi\approx\frac{\pi i}{\sqrt{2\epsilon}}+\pi,
\end{align}
whose imaginary part ($\epsilon$ dependent) exactly cancels $\Phi_{\rm cone}$, leading to
\begin{align}
\Phi_{\rm cone}+\Phi_{\rm lid}=\pi.
\end{align}
Similarly, one can obtain another $\pi$ flux from the lower cone, ending up with a total flux of $2\pi$.\\

\section{Example System: Antiferromagnetic Magnons}\label{supsec:2}

The PH Hamiltonian for antiferromagnetic magnons discussed in the main text is
\begin{align}
H_{\rm ph}=&\sigma_zH_0=t\sigma_z+ix\sigma_y-iy\sigma_x+a_0I,
\end{align}
whose eigenenergies are
\begin{align}
E_\pm=a_0\pm\sqrt{t^2-x^2-y^2}.
\end{align}
In terms of the hyperbolic parameterization
\begin{align}
t=\ell\cosh{\xi},\qquad
x=\ell\sinh{\xi}\cos{\lambda},\qquad
y=\ell\sinh{\xi}\sin{\lambda},
\label{eq:hyperbolic}
\end{align}
the right and left eigenvectors become
\begin{align}
\ket{\psi^{R}_-}=&
\begin{pmatrix}
-e^{-i\lambda}\sinh{\frac{\xi}{2}}\\
\cosh{\frac{\xi}{2}}
\end{pmatrix},\quad &
\ket{\psi^{R}_+}=
\begin{pmatrix}
e^{-i\lambda}\cosh{\frac{\xi}{2}}\\
-\sinh{\frac{\xi}{2}}
\end{pmatrix},\nonumber\\
\ket{\psi^{L}_-}=&
\begin{pmatrix}
e^{-i\lambda}\sinh{\frac{\xi}{2}}\\
\cosh{\frac{\xi}{2}}
\end{pmatrix},\quad &
\ket{\psi^{L}_+}=
\begin{pmatrix}
e^{-i\lambda}\cosh{\frac{\xi}{2}}\\
\sinh{\frac{\xi}{2}}
\end{pmatrix},
\label{eq:eigenvectors}
\end{align}
which are independent of $\ell$. Using Eqns.~(6)-(8) in the main text, we can obtain the Berry curvature for the lower band in the $(t, x, y)$-basis [\textit{i.e.}, $\mu$, $\nu$ running through $t$, $x$ and $y$]:
\begin{align}
F_{\mu\nu}=\frac{1}{2\ell^2}
\begin{pmatrix}
0 & -\sinh{\xi}\sin{\lambda} & \sinh{\xi}\cos{\lambda}\\
\sinh{\xi}\sin{\lambda} & 0 & -\cosh{\xi}\\
-\sinh{\xi}\cos{\lambda} & \cosh{\xi} & 0
\end{pmatrix},\label{eq:F-original}
\end{align}
which, in terms of the variable change Eq.~\eqref{eq:hyperbolic}, is identical to Eq.~(8) of the main text. Similarly, we can obtain the QGT for the lower band using
\begin{align}
Q_{\mu\nu}=&\bra{\partial_\mu\psi_-^L}(1-P)\ket{\partial_\nu\psi_-^R},
\end{align}
where $P=\ket{\psi_-^R}\bra{\psi_-^L}$. Regarding Eq.~(16) of the main text, we can then read off the quantum metric $g_{\mu\nu}$ from $Q_{\mu\nu}$ as
\begin{align}
g_{\mu\nu}=\frac{-1}{8\ell^2}
\begin{pmatrix}
2\sinh^2\xi & -\sinh(2\xi)\cos\lambda & -\sinh(2\xi)\sin\lambda \\
-\sinh(2\xi)\cos\lambda & 2+2\sinh^2\xi\cos^2\lambda & \sinh^2\xi\sin(2\lambda) \\
-\sinh(2\xi)\sin\lambda & \sinh^2\xi\sin(2\lambda) & 2+2\sinh^2\sin^2\lambda
\end{pmatrix},
\end{align}
which again, in terms of Eq.~\eqref{eq:hyperbolic}, is identical to Eq.~(17) of the main text.

Now, if we perform a coordinate transformation using Eq.~\eqref{eq:hyperbolic}, the Berry curvature and the quantum metric can be re-expressed in the $(\ell,\xi,\lambda)$-basis according to the tensor transformation
\begin{align}
 F_{\bar{\mu}\bar{\nu}}=F_{\mu\nu}\frac{\partial\mu}{\partial\bar{\mu}}\frac{\partial\nu}{\partial\bar{\nu}}, \qquad
 g_{\bar{\mu}\bar{\nu}}=g_{\mu\nu}\frac{\partial\mu}{\partial\bar{\mu}}\frac{\partial\nu}{\partial\bar{\nu}},
\end{align}
where $\bar{\mu}$, $\bar{\nu}$ refer to the new variables $(\ell,\xi,\lambda)$ while $\mu$, $\nu$ refer to the old variables $(t,x,y)$. Some straightforward algebra gives
\begin{align}
F_{\bar{\mu}\bar{\nu}}&=\frac12
\begin{pmatrix}
0 & 0 & 0\\
0 & 0 & -\sinh{\xi} \\
0& \sinh{\xi} & 0
\end{pmatrix}, \\
g_{\bar{\mu}\bar{\nu}}&=-\frac{1}4
\begin{pmatrix}
0 & 0 & 0\\
0 &1 & 0 \\
0 &0 & \sinh^2{\xi}
\end{pmatrix},
\end{align}
both of which are rank-2 tensors. This greatly simplified form is not a surprise because the eigenvectors in Eq.~\eqref{eq:eigenvectors}, hence the QGT, are independent of $\ell$. This means that the intrinsic geometry characterized by the QGT on all equal energy surfaces (\textit{i.e.}, arbitrary non-zero constant $\ell$) is identical, reducing the effective dimension of the parameter space to 2. Dropping the null rows and columns from the above matrices, we finally obtain the effective Berry curvature and the quantum metric residing in the $2$D parameter space spanned by $\xi$ and $\lambda$, justifying Eqs.~(23) and~(24) of the main text.

While the hyperbolic parameterization which eliminates the $\ell$-dependence in the eigenvectors (hence in the QGT) for all $2\times2$ PH models, the effective parameter space may not always be 2D because the eigenvectors should in general depend on $\theta$ and $\phi$ as well [the two angles specify the orientations of $\bm{n}_i$ in Eq.~(2) of the main text], whereas in the particular example of antiferromagnetic magnons we have set $\theta=0$ and $\phi=\pi/2$. On the other hand, in the $(t, x, y)$-basis the form of QGT is universal, as reflected by its independence of $\theta$ and $\phi$ under any conditions.

\end{document}